\documentclass[pra,aps,twocolumn,superscriptaddress]{revtex4-2}
\usepackage{graphicx,amsmath,color,soul}

\begin{document}

\title{Free-carrier-induced nonlinear dynamics in hybrid graphene-based photonic waveguides}

\author{Ambaresh Sahoo}
\affiliation{Department of Physics, Indian Institute of Technology Kharagpur, Kharagpur 721302, India}
\affiliation{Department of Physics, Indian Institute of Technology Guwahati, Guwahati 781039, India}

\author{Andrea Marini}
\affiliation{Department of Physical and Chemical Sciences, University of L'Aquila, Via Vetoio, 67100 L'Aquila, Italy}

\author{Samudra Roy}
\email{samudra.roy@phy.iitkgp.ac.in}
\affiliation{Department of Physics, Indian Institute of Technology Kharagpur, Kharagpur 721302, India}

\begin{abstract}
We develop from first principles a theoretical model for infrared pulse propagation in graphene-covered hybrid waveguides. 
We model electron dynamics in graphene by Bloch equations, enabling the derivation of the nonlinear conductivity and of a rate equation accounting for free-carrier generation. 
Radiation propagation is modeled through a generalized nonlinear Schr\"{o}dinger equation for the field envelope coupled with the rate equation accounting for the generation of free carriers in graphene. Our numerical simulations clearly indicate that unperturbed Kerr solitons accelerate due to the carrier-induced index change and experience a strong self-induced spectral blueshift. Our numerical results are fully explained by semianalytical predictions based on soliton perturbation theory.
\end{abstract}

\maketitle

\section{Introduction}
\vspace{-0.2cm}
Optical nonlinearity plays a key role in several modern and future photonic applications. In particular, nonlinear ultrafast effects at the femtosecond timescale are promising for optical information processing with petahertz bandwidth exceeding the speed of current electronic devices by six orders of magnitude \cite{Willner2014,Schoetz2019}. Guided waves in integrated nonlinear optical devices offer the fundamental advantage of radiation confinement leading to high field intensity that can be exploited over propagation \cite{Hendrickson2014}. Silicon photonics is promising for a plethora of integrated optical devices \cite{Thomson2016}, but the nonlinear optical functionalities of silicon are inherently limited by two-photon absorption \cite{Mizrahi1989,Tsang2002,Koos2007}. Plasmonic waveguides offer an alternative platform for integrated nonlinear optical processing \cite{Ozbay2006,Tuniz2020} thanks to the inherently high nonlinearity of metals \cite{Rotenberg2007,Scalora2010,Ginzburg2010,Marini2013,DeLeon2014,Tuniz2020bis} that is further enhanced by the surface confinement \cite{Marini2011,Skryabin2011,Li2017}. However, the high ohmic losses of metals pose a stringent limitation for the development of nonlinear plasmonic circuits \cite{Khurgin2015}. Absorption mitigation strategies in plasmonic devices resort to amplification schemes \cite{Bergman2003,Noginov2009,Marini2009,Stockman2010,DeLeon2010,Suh2012,Yang2015}, surface roughness reduction \cite{Wu2014}, and self-induced-transparency plasmon soliton excitation \cite{Marini2013bis}.

Graphene offers an appealing alternative to metals for integrated nonlinear optical applications \cite{Bonaccorso2010} thanks to ultrahigh electron mobility \cite{Novoselov2005,Wang2013} and its peculiar conical band structure \cite{CastroNeto2009}, which inherent anharmonicity leads to large nonlinear optical interaction \cite{Mikhailov2008,Wright2009,Ishikawa2010,Hendry2010,Wu2011,Cheng2014} that can be exploited, e.g., for high-harmonic generation \cite{Bowlan2014,Cox2017} and many other nonlinear optical applications \cite{Cox2019}. Furthermore, the ultrafast nonlinear dynamics of massless Dirac fermions (MDFs) in graphene  involve the photogeneration of free carriers (FCs) at the femtosecond timescale \cite{Baudisch2018}, which has already been employed for self-phase modulation in integrated nonlinear waveguides embedding graphene \cite{Vermeulen2018,Castello2020} and is promising for a plethora of future integrated optoelectronic applications.

Here, we investigate the ultrafast nonlinear dynamics of intense infrared radiation pulses propagating in a photonic waveguide covered by graphene, unveiling the role played by FCs in the spectral modulation observed. Within the MDF framework governing the graphene infrared response, we derive a generalized nonlinear Schr\"{o}dinger equation (GNLSE) accounting for the graphene Kerr nonlinearity, saturable absorption, and ultrafast dynamics of the photogenerated FCs. We further investigate soliton dynamics through a semianalytical technique based on Lagrange's variational method to theoretically capture the effect produced by FCs, deriving a set of ordinary differential equations describing the evolution of soliton parameters \cite{Bondeson1979,GPA}. Our calculations reveal that photogenerated FCs produce a peculiar signature in the temporal dynamics of optical solitons that undergo a substantial spectral blueshift accompanied by a temporal acceleration over propagation. Such a blueshift arises from the pulse self-action through a self-induced asymmetric refractive index temporal modulation produced by the photogenerated FCs. Our analytical predictions are confirmed by direct numerical simulations of the GNLSE and indicate that photonic waveguides covered by graphene offer a promising platform for integrated spectral modulation of infrared radiation.

\vspace{-0.3cm}

\section{Theoretical framework}
\begin{figure}[t]
\centering
\begin{center}
\includegraphics[width=0.48\textwidth]{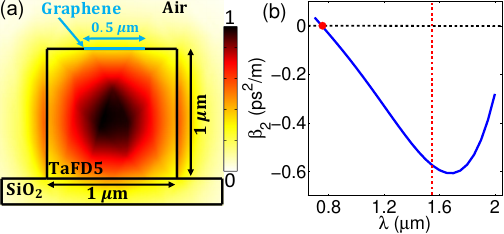}
\caption{(Color online) {\bf (a)} A TaFD5 glass waveguide with core dimensions (height $\times$ width) $1$ $\times \,1\,\mu$m covered by a single layer of graphene on top. The electric field intensity distribution $|{\bf e}(x,y)|^2$ of the fundamental quasi-TE mode at $\lambda_0=1.55 \ \mu$m is depicted in the inset in dimensionless units.  {\bf (b)} Group-velocity dispersion profile of the fundamental mode calculated by finite-element frequency-domain calculations \cite{COMSOL}. The input ($\lambda_0=1.55 \ \mu$m) and zero-dispersion ($\lambda_{\rm ZD} \simeq 0.75$ $\mu$m) wavelengths are indicated by the vertical dashed red line and the solid red circle, respectively.}
\label{GR_model}
\end{center}
\vspace{-0.4cm}
\end{figure}

\vspace{-0.2cm}
In order to describe the nonlinear ultrafast dynamics of infrared radiation pulses in hybrid graphene-covered waveguides, we consider a realistic design for a TaFD5 (dense tantalum flint) glass square waveguide covered by undoped graphene on top, deposited over a silica glass substrate and surrounded by air, as schematically depicted in Fig.\,\ref{GR_model}(a), which can be practically realized with existing state-of-the-art fabrication techniques. A single layer of graphene is deposited on top of the TaFD5 core where, upon infrared excitation, the FCs are accumulated over the propagation distance. The TaFD5 core is Raman inactive  \cite{Fedotova2006,Babushkin2007}, and linear and  nonlinear absorption of TaFD5 is negligible over approximately centimeter propagation distances. Hence TaFD5-based waveguides covered by single-layer graphene enable the full exploitation of photoexcited FCs in the graphene layer to attain efficient spectral modulation of the propagating pulse \cite{Christodoulides}. Alternatively, one can use silicon nitride ($\rm{Si_3N_4}$) as a core material with similar waveguide structure for the exact same purpose, as it is also Raman inactive and the linear and nonlinear losses are ultralow at the visible and infrared wavelengths \cite{Guo2018}.  In Fig.\,\ref{GR_model}(a) we illustrate a cross section of the waveguide and the intensity distribution of the fundamental quasi-transverse-electric (quasi-TE) mode at $\lambda_0 = 1.55$ $\mu$m that we calculate numerically by finite-element simulations \cite{COMSOL}. We further calculate numerically the group-velocity dispersion (GVD) of the waveguide quasi-TE mode in the $0.7-2.0$-$\mu$m wavelength range, which is indicated in Fig.\,\ref{GR_model}(b) by the solid blue curve.

\vspace{-0.5cm}
\subsection{Electron dynamics in graphene}
\vspace{-0.2cm}
To investigate the nonlinear spatiotemporal dynamics of infrared pulses in the proposed hybrid waveguide, we consider a pulse with electric field ${\cal E}({\bf r},t)={\rm Re}[u(z,t){\bf e}(x,y) {e}^{i\beta_0 z-i\omega_0 t}/\sqrt{P}]$, where $u(z,t)$ is the pulse envelope, ${\bf e}(x,y)$ and $P$ are the quasi-TE mode profile and scaling modal power calculated numerically, $\beta_0 \simeq 6.2967\times 10^6$\,rad$/$m is the carrier wave vector, and $\omega_0 = 2 \pi c/\lambda_0 \simeq 1.216\times10^{15}$\,rad$/$s is the carrier frequency. We emphasize that, in the considered geometry, the electric field profile of the quasi-TE mode ${\bf e}(x,y_0)\simeq {\rm e}_0\hat{\bf x}$ is practically uniform (independent of $x$) over the graphene layer at $y_0 = 1$ $\mu$m [see Fig. 1(a)] and directed over the $x$ direction. In turn, we approximate the electric field over the graphene flake as ${\bf E}(z,t)={\rm Re}[ {\bf E}_0(z,t) {e}^{i\beta_0 z-i\omega_0 t}]$, where ${\bf E}_0(z,t) = u(z,t){\rm e}_0\hat{\bf x}/\sqrt{P}$, which corresponds to the vector potential ${\bf A}(z,t)=-\int_{-\infty}^t {\bf E}(z,t')\,dt'$. In the MDF picture, the two-component spinor $\psi_{\bf k}$ of a single electron in graphene with initial in-plane wave vector ${\bf k}$ satisfies the single-particle time-dependent Dirac equation 
\begin{align} \label{eq1}
i\hbar\,\partial_{t}\psi_{\bf k}(z,t)=\hat{H}_{\bf k}(z,t)\psi_{\bf k}(z,t),
\end{align}
with spatiotemporal-dependent Hamiltonian $\hat{H}_{\bf k}(z,t)={\text{v}}_{\rm F}\,\mbox{\boldmath${\pi}$}\cdot\mbox{\boldmath${\sigma}$}$, where ${\text{v}}_{\rm F}\simeq c/300$ is the Fermi velocity, $\mbox{\boldmath${\pi}$}(z,t)=\hbar {\bf k} + e {\bf A}(z,t)$ is the electron quasi-momentum with $-e$ being the electron charge, and $\mbox{\boldmath${\sigma}$} = \left(\sigma_x,\sigma_y \right)$ is the two-dimensional Pauli-matrix vector. We emphasize that, owing to the conical band-structure of graphene, infrared excitation is resonant at $k_{\rm res} = \omega_0/2{\text{v}}_{\rm F} \simeq 6.1 \times 10^8$ rad$/$m and in turn $\beta_0 \ll k_{\rm res}$. Therefore we remark that the $z$ dependence of the Hamiltonian $\hat{H}_{\bf k}(z,t)$ is intended as adiabatic. Introducing the spatiotemporal-dependent energy $\varepsilon_{\bf k}(z,t) = {{\text{v}}_{\rm F}}\left|\mbox{\boldmath${\pi}$}(z,t)\right|$, one can express the Hamiltonian above as
\begin{align} \label{eq2}
\hat{H}_{\bf k}(z,t)=\varepsilon_{\bf k} (z,t)\left( \begin{matrix}
   0 & {{e}^{-i\theta_{\bf k} (z,t)}}  \\
   {{e}^{i\theta_{\bf k} (z,t)}} & 0  \\
\end{matrix} \right),
\end{align}
where $\theta_{\bf k}(z,t) = {\rm atan}\left[ \pi_y(z,t)/\pi_x(z,t) \right]$. Following a previously reported nonperturbative approach \cite{Marini2017}, we set the spinor ansatz as a linear combination of upper and lower states $\psi^{\pm}_{\bf k}$
\begin{align} \label{eq3}
\psi_{\bf k}(z,t) = c^+_{\bf k}(z,t)\psi^+_{\bf k}(z,t)+c^-_{\bf k}(z,t)\psi^-_{\bf k}(z,t),
\end{align}
where $\psi^\pm_{\bf k}(z,t) = (1/\sqrt{2}){e}^{\mp i\Omega_{\bf k} (z,t)}\left[{e}^{-i\theta_{\bf k}(z,t)/2}; \pm {e}^{i\theta_{\bf k}(z,t)/2} \right]^{\rm T}$ and $\Omega_{\bf k}(z,t)=(1/\hbar)\int_{-\infty}^{t}{\varepsilon_{\bf k}(z,t')d{t}'}$ is the instantaneous phase. Inserting the ansatz given by Eq.\,\eqref{eq3} into Eq.\,\eqref{eq1}, one gets the temporal evolution of the coefficients $c^\pm_{\bf k}(z,t)$,
\begin{align} \label{eq4}
\dot{c}^\pm_{\bf k}(z,t) = \frac{i}{2 }\dot{\theta}_{\bf k}(z,t)c^\mp_{\bf k}(z,t)\exp\left[\pm 2i\Omega_{\bf k}(z,t) \right].
\end{align}
Introducing the interband coherence $\rho_{\bf k} =c^+_{\bf k}c^{-*}_{\bf k}$ and population difference $n_{\bf k} = \left|c^+_{\bf k}\right|^2-\left|c^-_{\bf k}\right|^2$, one can rewrite the equation for the coefficients in the form of  Bloch equations, 
\begin{align} \label{eq5}
& \dot{\rho}_{\bf k} = -\gamma \rho_{\bf k}  -\frac{i}{2}\dot{\theta}_{\bf k} n_{\bf k}\exp(2i\Omega_{\bf k}), \\ 
& \dot{n}_{\bf k}    = -\gamma (n_{\bf k}+1) - i         \dot{\theta}_{\bf k} \exp(-2i\Omega_{\bf k})\rho + {\rm c.c.},  
\end{align}
where we have introduced the effective recombination rate $\gamma = (100\,{\rm fs})^{-1}$ accounting for electron-electron and electron-phonon collisions \cite{Baudisch2018}. We emphasize that electron-electron collisions in principle lead to a fast recombination time (tens of femtoseconds) \cite{Johannsen2013,Malic2012} leading to polarization dephasing. However, because we here focus on the effect of the photo-generated free carriers over optical propagation, we adopt an effective recombination time accounting only for electron-phonon collisions, which produce population recombination.

The induced single-electron current, in turn, is given by ${\bf j}^{\rm e}_{\bf k}(z,t) = - e \psi_{\bf k}^{\dagger}(z,t)\nabla_{\mbox{\boldmath${\pi}$}}H_{\bf k}(z,t)\psi_{\bf k}(z,t)$,  and explicitly
\begin{eqnarray} 
j^{\rm e}_{{\bf k},x}(z,t) & = & - e {\rm v}_{\rm F} \left[(n_{\bf k}(z,t)+1)\cos\theta_{\bf k}(z,t)  \right. \label{th1} \\
                           &   & \left. + i\sin\theta_{\bf k}(z,t) \left\{\rho_{\bf k}(z,t) \,  {e}^{-2i\Omega_{\bf k}(z,t)} - {\rm c.c.} \right\}  \right]. \nonumber
\end{eqnarray}
Then, the macroscopic current ${\bf J}(z,t)$ is calculated by integrating over all in-plane electron wave vectors 
\begin{align} \label{th2}
{\bf J}(z,t)=\frac{g_{\rm s}{g_{\rm v}}}{(2\pi)^2}\hat{\bf x}\int_{-\infty}^{+\infty}dk_x\int_{-\infty}^{+\infty}dk_y j^{\rm e}_{{\bf k},x}(z,t),
\end{align}
where $g_{\rm s} = g_{\rm v} = 2$ are the spin and valley degeneracy factors. Because even at high peak intensities of the order of TW$/$cm$^2$ $eA \ll \hbar k_{\rm res}$, in Eqs. (\ref{th1}) and (\ref{th2}) we neglect the optical momentum $e {\bf A}$ with respect to the electron momentum $\hbar {\bf k}$, obtaining 
\begin{eqnarray}
{\bf J}(z,t) & \simeq & {\rm Re} \left[ \frac{i e^2 {\rm v}_{\rm F}}{\pi^2\omega_0\hbar}K_{\rm FC}(z,t) {\bf E}_0(z,t){e}^{i\beta_0 z-i \omega_0 t}\right] \nonumber \\
             &   & + \frac{2e {\rm v}_{\rm F}}{\pi^2}\hat{\bf x} \int\limits_{0}^{\infty }{k\,dk}\int\limits_{0}^{2\pi }{d\phi }\sin\phi {\rm Im}\Gamma_{\bf k}(z,t) , \label{CurrentEq}
\end{eqnarray}
where $K_{\rm FC}(z,t) = ( 2 {\rm v}_{\rm F}/\omega )\int_0^\infty kdk\int_0^{2\pi}d\phi[ n_{\bf k}(z,t)+1]$ is the FC linear density, $\Gamma_{\bf k}(z,t)=\rho_{\bf k}(z,t){e}^{-2i\Omega_{0,k} t }$, and $\Omega_{0,k} = {\rm v}_{\rm F}k$. To unveil the role played by FCs, we first calculate the stationary solution of Eqs. (5) and (6) in the slowly varying envelope approximation by setting the ansatz $\Gamma_{\bf k} =\Gamma^{+}_{\bf k} e^{i\omega_0 t}+\Gamma^{-}_{\bf k} e^{-i\omega_0 t}$ and $n_{\bf k} = n_{0,{\bf k}}$, obtaining
\begin{subequations}
\begin{align}
& {\cal G}_{\bf k} = (\Gamma^{+}_{\bf k}-\Gamma^{-*}_{\bf k})/E_0^* = - iek_y n_{0,{\bf k}}\eta_{{\rm U},k}/2\hbar k^2\eta_{{\rm L},k}, \\
& n_{0,{\bf k}}                            = - \left[ 1+\frac{e^2k_y^2(\gamma^2+4\Omega_{0,k}^2+\omega_0^2)}{2\hbar^2k^4\eta_{{\rm L},k}}|E_0|^2 \right]^{-1}, \\
& \eta_{{\rm L},k} = [\gamma^2+(2\Omega_{0,k}+\omega_0)^2][\gamma^2+(2\Omega_{0,k}-\omega_0)^2], \\
& \eta_{{\rm U},k} = (\gamma +i\omega_0)[(\gamma^2+4\Omega_{0,k}^2-\omega_0^2)-2i\gamma\omega_0]. 
\end{align}
\end{subequations}
Thus, inserting the expressions above in Eq. (6) and integrating over the electron wave vectors both sides of Eq. (6), we obtain a rate equation for the FC density:
\begin{align}
{\partial_t}{K}_{\rm FC}(z,t) = -\gamma K_{\rm FC}(z,t) + \Upsilon(z,t) |E_0(z,t)|^2,
\end{align}
where $\Upsilon(z,t) = (\pi^2{\rm v}_{\rm F}e^2/2\hbar^2\omega_0^2)/\sqrt{1+|E_0(z,t)|^2/E_{\rm sat}^2}$ and $E_{\rm sat}=\sqrt{2I_{\rm S}/\epsilon_0 c} = 4.92 \times 10^6$ V$/$m with $I_{\rm S}$ being the saturable intensity of graphene. The equation above accounts for the photogeneration of FCs, which affects radiation dynamics in the hybrid waveguide through the induced macroscopic current ${\bf J}(z,t)$ in Eq. (\ref{CurrentEq}), which gives explicitly
\begin{align} 
{\bf J} = {\rm Re} \left\{\left[ \frac{i e^2 {\rm v}_{\rm F}}{\pi^2\omega_0\hbar}K_{\rm FC} + \sigma(|{\bf E}_0|^2)\right]{\bf E}_0 { e}^{i\beta_0 z-i\omega_0 t}\right\},  
\end{align}   
where $\sigma(|{\bf E}_0|^2)$ is the nonlinear conductivity at the carrier frequency $\omega_0$, provided by
\begin{align}
\sigma(|{\bf E}_0|^2) = \frac{2 i e {\rm v}_{\rm F}}{\pi^2} \int_0^\infty k dk \int_0^{2\pi} d\phi \sin \phi {\cal G}_{\bf k}^*(|E_0|^2).   
\end{align} 
Following the approach indicated in Ref. \cite{GRLREF}, we solve numerically the above integral for several radiation intensities $I = (1/2) \epsilon_0 c |E_0|^2$, finding excellent fitting with the analytical expression
$\sigma(I)$:
\begin{eqnarray}
\sigma (I) = \sigma_0 \left[ \frac{1}{ \sqrt{ 1 + I/I_{\rm S} } } - i \frac{ 1 - e^{ - \eta_1 \sqrt{ I/I_{\rm S} } } }{ \sqrt{1 + \eta_2 (I/I_{\rm S})^{0.4} } } \right], \nonumber
\end{eqnarray}
where $\sigma_0 = e^2/4\hbar$, $I_{\rm S} = 137 \hbar \omega_{\rm S}^2 \omega_0^2 / (8\pi {\rm v}_{\rm F}^2)$, $\omega_{\rm S} = 6.16$ rad/ps, $\eta_1 = \omega_\eta/\omega_0$, $\eta_2 = (\omega_\eta/\omega_0)^{0.8}$, and $\omega_\eta = 46.20$ rad/ps \cite{GRLREF}.

\vspace{-0.1cm}

\subsection{Radiation dynamics in the hybrid waveguide}
\vspace{-0.2cm}
In order to model radiation dynamics, we assume that nonlinear optical effects are sufficiently weak not to affect the mode structure of the hybrid graphene-covered waveguide. In turn, we calculate the quasi-TE mode profile by finite-element numerical simulations \cite{COMSOL} at $\lambda_0 = 1.55$ $\mu$m neglecting nonlinearity of graphene and the glass core. Note that, for the considered geometry, at $\lambda_{\rm ZD} \simeq 0.75$ $\mu$m the GVD coefficient ($\beta_2$) vanishes [see Fig. 1(b)], while at the operating wavelength ($\lambda_0=1.55 \ \mu$m) $\beta_2\simeq-0.576$ ps$^2$/m, the third-order dispersion coefficient is $\beta_3\simeq 0.599 \times 10^{-3}$ ps$^3$/m, and the Kerr coefficient of TaFD5 is $\gamma_{\rm TaFD5}\simeq 0.424$ W$^{-1}$m$^{-1}$. The scaling modal power $P$ at $\lambda_0$ is evaluated by the expression $P =\int_{\rm Full\,area}{ dx\,dy} \,{\rm Re} [{\bf e} \times {\bf h}^*]\cdot\hat{\bf z} \simeq 4.434 \times 10^{-12}$ W, where ${\bf e}(x,y)$ and ${\bf h}(x,y)$ are the numerically calculated electric and magnetic field profiles of the quasi-TE mode \cite{COMSOL}. We further evaluate the average $x$ component of the electric field amplitude experienced by the graphene layer at the top midpoint of the TaFD5 surface ${\rm e}_0\simeq (9.178\times10^{-4}+ i\,16.288)$ V/m. Note that at $\lambda_0$ the GVD coefficient is negative, $\beta_2<0$, thus enabling the excitation of bright optical solitons. 

Performing an asymptotic expansion of Maxwell's equations, where the nonlinearity of graphene and of the waveguide core are treated as perturbations, following a standard approach detailed in previous works  (see, e.g., Ref. \cite{{Andriy}}), we derive a generalized nonlinear Schr\"{o}dinger equation (GNLSE) for the dimensionless pulse envelope $\psi(\xi,\tau)=u/\sqrt{P_0}$ in the comoving reference frame 
\begin{align} \label{NLSE_norm}
&i\partial_{\xi} \psi+\hat{D}(i\partial_{\tau})\psi+ N^2\left(1+i\tau_{sh}\partial_\tau\right)|\psi|^2\psi -d_{\rm FC}\Phi_{\rm FC}\,\psi 
 \nonumber\\ &+i\frac{\alpha_{\rm 0} \psi}{\sqrt{1+3|\psi|^2/\psi_{\rm sat}^2}}+ \frac{\alpha_{\rm 0} \psi \left(1-e^{-\eta_1\sqrt{3|\psi|^2/\psi_{\rm sat}^2}}\right)}{\sqrt{1+\eta_2\left(3|\psi|^2/\psi_{\rm sat}^2 \right)^{0.4}}}=0, 
\end{align} 
where the dimensionless variable $\xi=z/L_D$ represents the longitudinal coordinate $z$ rescaled to the dispersion length $L_{\rm D} = t_0^2/|\beta_2|$ and $\tau = (t-z/v_{\rm g})/t_0$ represents the  temporal coordinate $t$ in the comoving reference frame shifted by $z/v_{\rm g}$ and rescaled to the pulse duration $t_0$, while $v_{\rm g} = 1.5133\times 10^8$ m$/$s is the group velocity, $\hat{D}(i\partial_{\tau})$ is the normalized dispersion operator, $\alpha_{\rm 0}=w |{\rm e}_0|^2 \sigma_0 L_{\rm D} / P$ is the normalized linear absorption rate with $w$ being the width of the graphene layer, $L_{\rm NL}=1/(\gamma_{\rm TaFD5}P_0)$ is the nonlinear length with $P_0$ being the input peak power, $N^2=L_D/L_{\rm NL}$ is the soliton order, and $\tau_{sh}=1/\omega_0 t_0$ is the normalized self-steepening coefficient. As anticipated in the previous section, the GNLSE is coupled through the coupling coefficient $d_{\rm FC}=e^2{\rm v}_{\rm F} w |{\rm e}_0|^2/\pi^2\hbar\omega_0 P= 6.0678\times10^{-7}$, with the rate equation governing the temporal evolution of the normalized FC-density spatiotemporal profile $\Phi_{\rm FC}(\xi,\tau) = L_{\rm D} K_{\rm FC}(z,t)$, explicitly given by 
\begin{align} \label{FC_norm}
\frac{\partial {{\Phi}_{\rm FC}}}{\partial\tau} = \frac{\Theta_{\rm FC}|\psi|^2}{\sqrt{1+3|\psi|^2/\psi_{\rm sat}^2}}-\frac{\Phi_{\rm FC}}{\tau_{\rm FC}},
\end{align} 
where $\Theta_{\rm FC}=L_{\rm D} t_0 P_0 |{\rm e}_0|^2 \pi^2{\rm v}_{\rm F} e^2/(2\hbar^2P\omega_0^2)$ is the normalized FC generation rate, $\psi_{\rm sat}^2 = 3|E_{\rm sat}/{\rm e}_0|^2P/P_0$ is the normalized saturation intensity, and $\tau_{\rm FC}=1/\gamma t_0$ is the normalized FC recombination rate. 

\begin{figure}[t]
\centering
\begin{center}
\includegraphics[width=0.48\textwidth]{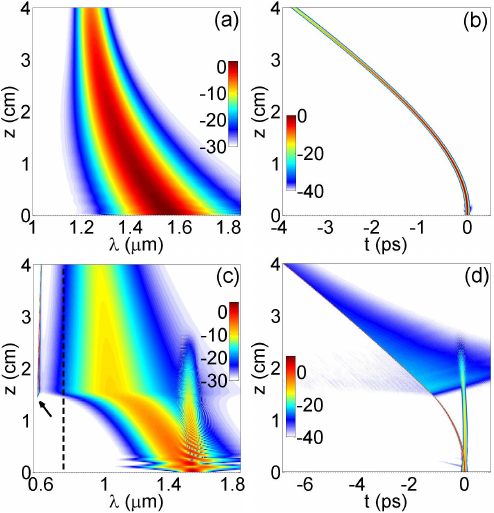}
\caption{(Color online) Longitudinal evolution of (a) and (c) spectral ($|\widetilde{\psi}(\xi,\omega)|^2$) and (b) and (d) temporal ($|\psi(\xi,\tau)|^2$) dynamics of input pulses $\psi(0,\tau) = {\rm sech}(\tau)$ with distinct soliton orders (a) and (b) $N=1$ ($P_0= 1.358\times 10^4$ W, $t_0 = 10$\,fs, $\alpha_0=0.316$, $\Theta_{\rm FC}=1.0865\times 10^7$, $\psi_{\rm sat}^2=8.966\times 10^{-5}$) and (c) and (d) $N=2$ ($P_0= 8.692\times 10^3$ W, $t_0 = 25$\,fs, $\alpha_0=1.975$, $\Theta_{\rm FC}=1.0865\times 10^8$, $\psi_{\rm sat}^2=1.401\times 10^{-4}$). The vertical dashed line in (c) locates the zero-dispersion wavelength $\lambda_{\rm ZD}$ across which resonant radiation (indicated by the arrow) arises due to suppression of FC-mediated self-frequency blueshift \cite{Skryabin2003}.}
\label{GR_blueshift}
\end{center}
\vspace{-0.5cm}
\end{figure}

\vspace{-0.2cm}
\section{FC-induced spectral dynamics}
\vspace{-0.2cm}

\begin{figure}[t]
\centering
\begin{center}
\includegraphics[width=0.48\textwidth]{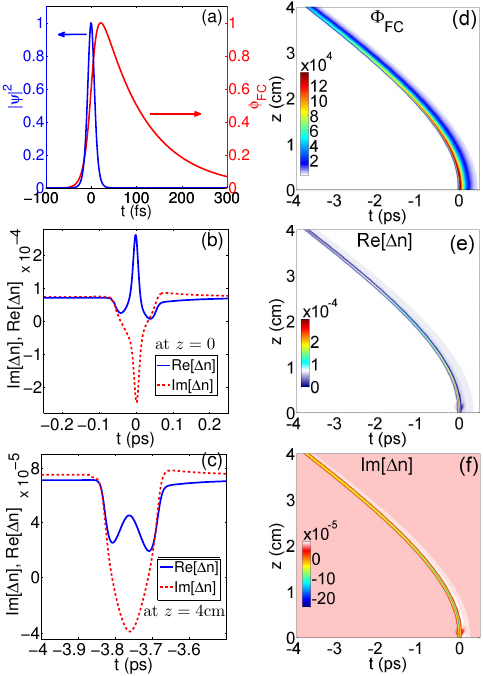}
\caption{(Color online) (a) Normalized temporal profile of the  FC density $\Phi_{\rm FC}(z/L_{\rm D},t/t_0)$ (red curve corresponding to the right vertical axis) at $z=0$, and temporal profile of the impinging pulse intensity $|\psi(0,t/t_0)|^2$ (blue curve corresponding to the left vertical axis) for $t_0=10$\,fs. (d) Spatiotemporal evolution of $\Phi_{\rm FC}(z/L_{\rm D},t/t_0)$.  Real and imaginary parts of the FC-induced refractive index modulation $\Delta n (z/L_{\rm D},t/t_0)$ [see Eq.\,\eqref{del_n}] at (b) input $z=0$ and (c) output $z=4$ cm, and their respective spatiotemporal evolutions (e) and (f) for $N=1$.}  \label{dynamicsFC}
\end{center}
\vspace{-0.5cm}
\end{figure}

The propagation of infrared radiation pulses with normalized envelope $\psi$ in the considered graphene-covered hybrid waveguide is governed by Eqs. (\ref{NLSE_norm}) and (\ref{FC_norm}) that, as explained above, account for the nonlinear dynamics produced by the photogenerated FCs in graphene. By adopting the standard split-step fast Fourier transform complemented with the fourth-order Runge-Kutta method, we solve Eqs.\,(\ref{NLSE_norm}) and (\ref{FC_norm}) numerically considering input pulses in the form of Kerr solitons, $\psi(0,\tau)={\rm sech}(\tau)$. In Fig. \ref{GR_blueshift} we illustrate the spectral and temporal evolutions of fundamental ($N=1$) [Figs. \ref{GR_blueshift}(a) and \ref{GR_blueshift}(b)] and second-order ($N=2$) [Figs. \ref{GR_blueshift}(c) and \ref{GR_blueshift}(d)] solitons. Note that, owing to FC excitation in the graphene layer, higher-order nonlinearity and dispersion, and absorption, the spatiotemporal evolution of Kerr solitons in the hybrid waveguide leads to a strong self-induced spectral blueshift $\Delta\lambda \simeq 250$ nm [see Fig. \ref{GR_blueshift}(a)], as a result of acceleration dynamics in the time domain [see Fig. 2(b)], over a propagation distance of $4$ cm. The main limiting factor to obtain larger spectral modulation is represented by graphene absorption, which quenches the pulse propagation and limits frequency conversion efficiency. However, for higher-order solitons with increased power, graphene absorption becomes saturated as a consequence of partial Pauli blocking produced by the photoexcited FCs, and the self-induced spectral blueshift increases [see Fig. \ref{GR_blueshift}(c) for soliton order $N=2$]. Note in Fig. \ref{GR_blueshift}(c) that, thanks to absorption saturation, the self-induced spectral blueshift becomes as large as $\Delta\lambda \simeq 600$ nm, thus enabling a large spectral tunability of the propagating pulse.

To understand the underlying physics behind FC-induced spectral modulation, in Fig.\,\ref{dynamicsFC} we plot the normalized FC density  $\Phi_{\rm FC}(0,t/t_0)$ at the waveguide input [Fig.\,\ref{dynamicsFC}(a)] and $\Phi_{\rm FC}(z/L_{\rm D},t/t_0)$ as a function of the propagation distance $z$ [Fig.\,\ref{dynamicsFC}(d)]  when the waveguide is excited by a fundamental Kerr soliton ($N=1$), whose intensity profile $|\psi(0,t/t_0)|^2 = {\rm sech}^2(t/t_0)$ is indicated by the blue curve in Fig. \ref{dynamicsFC}(a). Owing to the asymmetric temporal profile of $\Phi_{\rm FC}(z/L_{\rm D},t/t_0)$, the leading and trailing edges of the pulse experience different FC-induced modulation, which produces temporal acceleration of the input soliton accompanied by a self-induced frequency blueshift. We evaluate the effect of FC-induced modulation by calculating the correction to the refractive index  
\begin{align} \label{del_n}
&\Delta n(z/L_{\rm D},t/t_0) =\sqrt{\epsilon_{\rm L}+\epsilon_{\rm NL}}-\sqrt{\epsilon_{\rm L}},
\end{align}
where $\epsilon_{\rm L}=1+{i\sigma_0}/{(\omega_0\epsilon_0 t_{\rm gr})}$ with $t_{\rm gr}$\,($=0.3$\,nm) being the thickness of the monolayer graphene and 
\begin{align}
\epsilon_{\rm NL}&=\frac{2c}{\omega_0 L_{\rm D}}\left[N^2(1+i\tau_{sh}\partial_\tau)|\psi|^2 \right. \nonumber\\
&\left.-d_{\rm FC}\Phi_{\rm FC} +i{\alpha_{\rm 0}}/\sqrt{1+3|\psi|^2/\psi_{\rm sat}^2} \right. \nonumber \\
&\left.+\alpha_{\rm 0}\left(1 -e^{-\eta_1\sqrt{3|\psi|^2/\psi_{\rm sat}^2} }\right)/\sqrt{{1+\eta_2\left(3|\psi|^2/\psi_{\rm sat}^2 \right)^{0.4}} } \right] \nonumber
\end{align} 
are the linear and nonlinear dielectric constants, respectively, whose  real and  imaginary parts at the input ($z=0$) [Fig.\,\ref{dynamicsFC}(b)] and output ($z=4$cm) [Fig.\,\ref{dynamicsFC}(c)] are illustrated. The spatiotemporal dynamics of the real [Fig.\,\ref{dynamicsFC}(e)] and imaginary [Fig.\,\ref{dynamicsFC}(f)] parts of the modulated refractive index $\Delta n$  are shown. Such a time-dependent and asymmetric (solely due to the graphene FCs) refractive index modulation is the inherent ingredient producing a time-dependent phase shift leading to the strong spectral modulation illustrated in Fig. \ref{GR_blueshift}. 

In order to shed further light on pulse propagation dynamics, we develop a variational treatment where we recast Eq.\,\eqref{NLSE_norm} (after neglecting higher-order terms) in the form of a perturbed NLSE \cite{GPA}: $i\partial_\xi \psi+\frac{1}{2}\partial^2_\tau \psi+|\psi|^2\psi=i\epsilon(\psi)$, where 
\begin{eqnarray}
& & \epsilon (\psi)  =  -id_{\rm FC}\Phi_{\rm FC}\psi -\alpha_{\rm 0} \psi/{\sqrt{1+3|\psi|^2/\psi_{\rm sat}^2}} \\
& & + i\alpha_{\rm 0} \psi \left(1-{e}^{-\eta_1\sqrt{3|\psi|^2/\psi_{\rm sat}^2}}\right)/\sqrt{1+\eta_2(3|\psi|^2/\psi_{\rm sat}^2)^{0.4}}. \nonumber 
\end{eqnarray}
A Lagrangian density ${\cal L}_{\rm D}$ for such a system can be defined as ${\cal L}_{\rm D} = (i/2) ( \psi^* \partial_\xi \psi - \psi \partial_\xi \psi^*)  + (1/2)(|\psi|^4 - |\partial_\tau \psi|^2) -2{\rm Re}[i\epsilon u^*]$. The total Lagrangian $L$ is calculated from the Lagrangian density as $L=\int_{-\infty}^{\infty}\mathcal{L_D}\,d\tau$. Adopting the ansatz $\psi= \sqrt{E\eta/2} {\cal F}(\tau){e}^{i\phi-i\Omega_{\rm p}(\tau-\tau_{\rm p})-i\beta(\tau-\tau_{\rm p})^2}$, where ${\cal F}(\tau)={\rm sech}{[\eta(\tau-\tau_{\rm p})]}$ and the six dimensionless parameters $-$ energy $E$, pulse width $\tau_{\rm w}=2/\eta$, temporal position $\tau_{\rm p}$, phase $\phi$, frequency shift $\Omega_{\rm p}$, and chirp $\beta$ $-$ become adiabatically evolving functions of the propagation distance $\xi$. The Ritz optimization procedure \cite{Anderson} for $L$  leads to a set of coupled ordinary differential equations (ODEs) describing the evolution of individual pulse parameters
\begin{subequations}\label{SolPerThEqs}
\begin{align}
& \frac{dE}{d\xi } = -2\alpha_{\rm 0}E \, {\cal K} \,\ {\rm atan}(1/{\cal K}), \label{var1}\\ 
& \frac{d\tau_{\rm p}}{d\xi} =-\Omega_p,  \label{var2}\\
& \frac{d\Omega_{ p} }{d\xi } = d_{\rm FC}\Theta_{\rm FC}(E\eta^3/2)\int_{-\infty}^{\infty}d\tau\, {e}^{-\tau/\tau_{\rm FC}}  \label{var3} \\
& \hspace{10mm} \times {\cal S}(\tau) {\cal F}^2(\tau) \int_{-\infty}^{\tau}\frac{{\cal F}^2(\tau'){e}^{\tau'/\tau_{\rm FC}} }{\sqrt{1+ {\cal F}^2(\tau')/{\cal K}^2}} d\tau', \nonumber \\ 
& \frac{d\eta }{d\xi } = - \alpha_{\rm 0} \eta \, {\cal K} \,\ {\rm atan}(1/{\cal K}) + 2 \beta \eta  \label{var4} \\ 
& \hspace{10mm}  + (6/\pi^2)\alpha_{\rm 0} \eta^4 \int_{-\infty}^{\infty} \frac{(\tau-\tau_{\rm p})^2 {\cal F}^2(\tau)}{\sqrt{1 + {\cal F}^2(\tau)/{\cal K}^2}} d\tau, \nonumber \\
& \frac{d\beta}{d\xi } = 2\beta^2 + \frac{1}{\pi^2} (E\eta^3 - 2\eta^4) + \int_{-\infty}^{\infty} d\tau {\cal F}^2(\tau)[1-2{\cal P}(\tau)]   \nonumber \\
&\hspace{10mm} \times \frac{3}{2\pi^2} \left\{  2\alpha_{\rm 0} \eta^3 \frac{1- {e}^{-\eta_1{\cal F}(\tau)/{\cal K}}}{\sqrt{1 + \eta_2  {\cal F}^{0.8}(\tau)/{\cal K}^{0.8} } }  \right. \label{var5} \\
&\hspace{10mm} \left. - d_{\rm FC}\Theta_{\rm FC} E\eta^4 {e}^{-\tau/\tau_{\rm FC}} \int_{-\infty}^{\tau}\frac{{\cal F}^2(\tau'){ e}^{\tau'/\tau_{\rm FC}} d\tau'}{\sqrt{1 + {\cal F}^2(\tau')/{\cal K}^2}} \right\} , \nonumber 
\end{align}
\end{subequations}
where ${\cal K}=\sqrt{2\psi_{\rm sat}^2/3E\eta}$, ${\cal S}(\tau) = {\rm tanh}[\eta(\tau-\tau_{\rm p})]$ and ${\cal P}(\tau) = \eta(\tau-\tau_{\rm p}){\cal S}(\tau)$.
  
\begin{figure}[t]
\centering
\begin{center}
\includegraphics[width=0.48\textwidth]{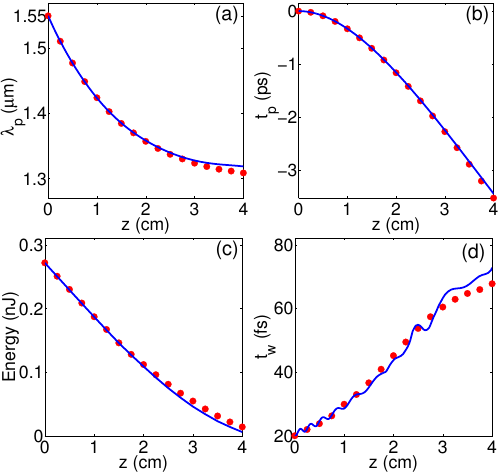}
\caption{(Color online) Evolution over the longitudinal coordinate $z$ of (a) pulse central wavelength $\lambda_{\rm p}=2\pi c/(\omega_0 + \Omega_{\rm p}/t_0)$, (b) temporal position $t_p=\tau_{\rm p}t_0$, (c) pulse energy $E$, and (d) pulse width $t_{\rm w}=\tau_{\rm w}t_0$ of the perturbed Kerr soliton for $N=1$. The solid blue curves indicate predictions from the variational approach through the solution of the coupled ODEs, whereas red circles represent numerical data obtained from the full numerical integration of Eqs. (14) and (15) by the split-step fast Fourier transform algorithm complemented with the fourth-order Runge-Kutta method (neglecting the higher-order terms).}  
\label{GR_blueshift_var}
\end{center}
\vspace{-0.5cm}
\end{figure}

This set of coupled ODEs provides significant physical insight into the role played by photoexcited FCs because they indicate how perturbations affect the individual pulse parameters. Note, in particular, that photoexcited FCs in graphene directly affect the soliton carrier frequency $\Omega_{\rm p}$ because the normalized FC  generation rate $\Theta_{\rm FC} $ appears on the right-hand side of Eq. (\ref{var3}) with an overall positive sign. This immediately implies that soliton dynamics in the hybrid waveguide system drifts towards lower central wavelength accompanied by a temporal acceleration that can be directly inferred from Eq.\,\eqref{var2}. The FC-generation-induced absorption $\alpha_0$ enters Eq. (\ref{var1}) for the soliton energy $E$, producing energy quenching. The dynamics of the soliton temporal width $\tau_{\rm w}$ and chirp $\beta$ is more involved and can be grasped only through numerical integration. We emphasize that the equation for phase $\phi$ is overlooked here because it does not affect any other soliton parameter. In order to validate the variational predictions, in Fig. \ref{GR_blueshift_var} we compare the results from the variational approach (blue curves) with data from full numerical simulation of Eqs. (14) and (15) by the split-step fast Fourier transform algorithm complemented with the fourth-order Runge-Kutta method for the fundamental soliton, $N=1$ (solid circles). The set of coupled ODEs [Eqs.\,\eqref{var1}$-$\eqref{var5}] is solved by numerical integration considering input pulse parameters as initial conditions. We find that results from the variational approach are in excellent agreement with direct numerical simulations of Eqs.(14) and (15). Both approaches indicate a strong central frequency blueshift up to $\Delta \lambda \simeq 250$\,nm for the fundamental soliton as a consequence of FC generation in the graphene layer [see Fig.\,\ref{GR_blueshift_var}(a)] and temporal acceleration [see Fig.\,\ref{GR_blueshift_var}(b)]. We finally note that the dissipative radiation dynamics ensuing from FC generation produces quenching of the pulse energy  [see Fig.\,\ref{GR_blueshift_var}(c)] and a broadening of the pulse width [see Fig.\,\ref{GR_blueshift_var}(d)]. However, as discussed above, absorption can be saturated by higher-energy pulses thanks to partial Pauli blocking of the photo-induced FCs (see Fig. 2), leading to self-induced spectral blueshift as large as $\Delta\lambda \simeq 600$ nm.

\section{Conclusions}

We have developed from first principles a theoretical model enabling the description of infrared pulse propagation in graphene-covered hybrid waveguides. Our theoretical model finely accounts for the waveguide dispersion, the nonlinearity of the waveguide core, and the effect produced by photogenerated free carriers in the graphene layer. Specializing our calculations to a realistic rectangular waveguide composed of TaFD5 glass on top of silica substrate and covered with graphene on top, we find that free-carrier generation enables efficient spectral modulation of ultrafast pulses propagating in the waveguide. In particular, we predict a strong spectral blueshift of several hundreds of nanometers for peak pulse powers of the order of $10$\,kW. We further validate numerical results with a variational approach indicating that the refractive index temporal modulation ensuing from free-carrier generation is responsible for the observed spectral modulation. Our results indicate that graphene-covered hybrid waveguides offer an appealing platform for frequency conversion in integrated optoelectronic devices.

\section*{ACKNOWLEDGMENT}
S.R. acknowledges funding from Shri Gopal Rajgarhia International
Programme$-$Indian Institute of Technology (SGRIP-IIT) Kharagpur for the collaborative visit of A.M.

\end{document}